\documentstyle[preprint,prl,aps]{revtex}
\draft
\title{Quantum soliton generation using an interferometer}
\author{M.~J. Werner$^*$}
\address{NTT Basic Research Laboratories, 3-1 Morinosato-Wakamiya, Atsugi-shi, Kanagawa-ken, 243-0198, Japan.}
\date{Received 11 March 1998}
\begin{document}  
\maketitle
\begin{abstract} 
For the first time a method for realizing macroscopic quantum optical solitons is presented.
Simultaneous photon-number and momentum squeezing is predicted using soliton propagation in an interferometer. 
Extraction of soliton pulses closer to true quantum solitons
than their coherent counterparts from mode-locked lasers is possible. 
Moreover, it is a general method of reducing photon-number fluctuations below the shot-noise level for non-soliton pulses
as well. It is anticipated that similar reductions in particle fluctuations 
could occur for other forms of interfering bosonic fields
whenever self-interaction nonlinearities exist, for example, interacting ultracold atoms.
\end{abstract}
\narrowtext
Quantum solitons are energy eigenstates
or photonic number states for optical systems. 
The particle number fluctuations of these quantum states are zero and this property is preserved by the
nonlinear system. A quantum soliton is
a fundamental state in nature capable of allowing information bits to propagate arbitrarily
long distances in the presence of dispersion and nonlinearity.
Even approximate forms of such an idealised quantum object have been difficult to observe.

Mode-locked lasers do not produce quantum solitons but instead have a Poissonian distribution of number states~\cite{nature}.
Solitons excited by such laser pulses have initial fluctuations in the four soliton parameters
of photon number, momentum, position and phase. Neither are they minimum uncertainty states.
Position fluctuations grow quadratically for a freely propagating fundamental soliton
due to the linear dispersion acting on the initial momentum fluctuations. Similarly phase noise grows
due to the initial fluctuations in its conjugate variable photon number acted upon by the nonlinearity.
Typically initial fluctuations in position and phase become insignificant compared to the increase due to
quantum diffusion.

In this Letter quantum soliton generation using an approach based on interference of optical 
fields is proposed. 
The idea that one can manipulate the internal quantum noise structure of propagating initially coherent solitons
to produce more than 11dB photon number squeezing using an asymmetric Sagnac loop 
was first put forward by the author recently~\cite{osa97}.
That followed surprising results on interference 
of solitonic fields in nonlinear optical loop mirrors 
to produce more than 15dB excess noise~\cite{cleo97}.
Here it is shown that in the easily accessible regime of macroscopic photon numbers of order $10^8-10^9$ typical
in picosecond and subpicosecond laser experiments, it is possible to produce soliton pulses with greater than 10dB  
reduction in photon number fluctuations. 
In addition, momentum fluctuations can be reduced to more than 6dB below that of a coherent state pulse.
Initial experiments have recently observed 3.9dB (6.0dB inferred from 79\% detection efficiency)~\cite{schmitt} 
(5.7 dB~\cite{krylov} has independently been observed) photon-number 
squeezing at room-temperature using the method disclosed earlier~\cite{osa97} and discussed here in more detail.

Any scheme which involves interference of bosonic fields
(different in either direction or polarization) and 
where at least one has undergone evolution according to an effective 
($1+1$)D nonlinear Schr\"odinger equation could exhibit similar behaviour.
This includes waveguided atomic solitons~\cite{zhangws:94},
spatial optical solitons for nonlinearities that depend on the particle
density flux and optical pulses in cascaded 
quadratic media or phase-mismatched $\chi^{(2)}$ simulton interactions~\cite{josab}.
Results are presented for an ideal Mach-Zehnder interferometer since the
two counterpropagating fields of a Sagnac fiber loop 
are assumed to have propagated in two independent waveguides.
It is important that the system is described using a dispersive quantum field theory in order to 
contain the essential physics and to make predictions for current optical pulse experiments 
which use pulse durations corresponding to $70\mbox{fs}<t_0<2\mbox{ps}$. 
For optical systems the nonlinearity and dispersion can be turned on/off in one arm easily.
For systems containing real massive interacting particles the loop results are more appropriate.
Optical experiments were performed using Sagnac loops to reduce low-frequency noise.
Importantly, the scheme is not critically sensitive to exact coupling ratios, powers or fibre lengths
although optimization is required to reach beyond the 10dB limit. 
The interference of optical fields has several distinct advantages 
over direct spectral filtering~\cite{friberg} including 
greater noise reduction, smooth output pulse envelopes and simultaneous photon-number/momentum squeezing.
The model investigated to demonstrate the idea -- the nonlinear Schr\"odinger equation -- is ubiquitious
in the physics of dispersive self-interacting fields and describes perturbations in a wide variety of physical systems.
It is anticipated that particle number fluctuations in more general systems such as described
by a quantum Ginzburg-Landau equation could exhibit similar behaviour to the idealized solitons
discussed in this Letter. 
There are strong reasons to believe this is the case especially for weak damping
which occurs in optical fibers in the $1.5\mu$m regime.

The nonlinear Schr\"odinger equation has been used in various forms 
for the study of Bose-Einstein condensation~\cite{ruprechthb95}, waveguided ultracold atoms~\cite{zhangws:94} and 
propagating  coherent quantum optical solitons~\cite{firstprl,laih:89,werner,kaertner} where
the quantum nonlinear Schr\"odinger equation (QNLSE) governs dynamics of the photon flux amplitude. 
The Raman-modified stochastic nonlinear Schr\"odinger equation 
for the normalised photon flux fields $\{\phi(\zeta,\tau),\phi^\dagger(\zeta,\tau)\}$ in the positive-P representation
is given by~\cite{carterd91}
\FL\begin{eqnarray}
{{\partial\ln\phi}\over{\partial\zeta}} = 
&-&\frac{i}{2}\bigl(1\pm{{\partial^2}\over{\partial\tau^2}}\bigr)
+if\phi^\dagger\phi+\sqrt{i}\Gamma_e \nonumber\\
&+&i\int_{-\infty}^\tau d\tau^\prime 
h(\tau-\tau^\prime)\phi^\dagger(\tau^\prime)\phi(\tau^\prime)+i\Gamma_v \ ,
\end{eqnarray}  
where length and time variables $(\zeta,\tau)$ in the comoving frame 
at speed $\omega^\prime$ (group-velocity at the carrier frequency) in the laboratory
frame $(x,t)$ are $\tau=(t-x/{\omega^\prime})/t_0,\zeta=x/x_0,x_0=t_0^2/|k^{\prime\prime}|$.
For this equation and its Hermitean conjugate for $\phi^\dagger$,
the characteristic time scale $t_0$ will be chosen to be the pulse width 
and the soliton period is $\pi/2$ times longer than
the dispersion length $x_0$ determined by $t_0$ and the second-order dispersion $k^{\prime\prime}$.
The quantum noise from the electronic nonlinearity $\Gamma_e$ is a real delta-correlated
Gaussian noise with variance given by the product of the electronic fraction 
$f$ (ideal QNLSE has $f=1,h(\tau)=\Gamma_v=0$) and 
inverse photon number scale $1/\bar n=\chi t_0/|k^{\prime\prime}|{\omega^\prime}^2$. 
Silica fiber Raman gain has a peak near $13$ THz and $f=0.81$ for
Raman inhomogeneous model parameters used 
in evaluating the response function $h(\tau)$ and noise $\Gamma_v$
corresponding to the Raman gain curve in Reference~\cite{carterd91}.
Quantum field propagation is performed numerically using techniques discussed elsewhere~\cite{wernerd}.
All simulations without Raman used $\bar n=10^8$ with averaging over $10^5$ trajectories
for the positive-P representation. 
Error bars in the plots represent the estimated combined sampling and step-size error. 

To illustrate the interference at the output beamsplitter of unequal amplitude solitonic fields consider
the interference term of the transmitted photon spectral flux proportional to
\begin{equation}
\phi_1^\dagger(-\omega)\phi_2(\omega)-\phi_2^\dagger(-\omega)\phi_1(\omega) \ .
\end{equation}
The interference term above is the same as obtained from a generalised quadrature-phase operator
measurement in homodyne detection where the local oscillator (weak field) can have 
nonclassical statistics~\cite{wernerchi2}. One can consider the beamsplitter
of a loop to act first as a state preparation device and later, combined with the weak field and
photon detector, as a measurement apparatus. It is easy to show that placing a phase-shifter
in the weak field arm can be used to switch from sub-Poissonian to super-Poissonian statistics
in analagy with changing a quadrature-phase measurement from the squeezed quadrature 
to the anti-squeezed quadrature.
Since the two fields $\phi_1$ and $\phi_2$ in each arm of the interferometer
typically has a different spectral distribution due to the different initial amplitudes,
there exists a spectral filtering mechanism. 
Importantly, it has been shown previously that the internal quantum noise
structure for the two fields will be quite different~\cite{cleo97,wernerf97}. 
These two effects combined will alter the quantum statistics of the resultant field. 
In addition, the interferometer is obviously sensitive to 
pulse chirp, a frequency dependent phase-shift, induced 
by group-velocity dispersion which is important for fiber lengths of a few soliton periods discussed here. 
These aspects are not present in any single-mode description as used in Reference~\cite{ritzeb79} 
where single polarization and nonlinear polarization rotation interferometric
configurations were considered with one arm as free space.
This paper goes beyond the one arm configuration and shows that noise reduction is possible even
for solitonic fields in both arms.

While a heuristic analysis based on the energy input-output curve 
for Sagnac loops can sometimes
estimate whether the output photon-number noise is expected to be below shot noise, it is generally not reliable.
It is suggested that an appropriate method for large photon numbers
which does not require the more sophisticated techniques used
with the positive-P representation is to use the truncated Wigner representation~\cite{drummondh:93}
which for the ideal QNLSE with coherent state inputs involves solving only the classical equations 
argmented with Gaussian noise on the initial conditions~\cite{wernerd}.
The classical nonlinear phase shift and dispersion in each arm of the loop,
which has long been known to support effective soliton switching~\cite{nolm} 
and reduction of dispersive waves~\cite{matsumotoih:94},
leads to a characteristic input-output curve for the highly asymmetric (90:10) and near balanced (60:40) case.
The transmitted pulse photon-number (scaled by $\bar n$) versus 
the input amplitude $N$ with $\phi(0,\tau)=N\mbox{sech}(\tau)$ inputs
is given in Fig.~\ref{figio}
for the case 90:10 at $\zeta=\pi,2\pi$ and 60:40 at $\zeta=2\pi$. 
A propagation distance longer than $\zeta=\pi$ for the 60:40 case 
was chosen so that input-output curves
contain at least one turning point.
One can see that after propagating 2 soliton periods for the 90:10 case,
the input-output curve's slope is positive leading to excess noise
at the output for all inputs except near the turning 
points $N=1.35,1.5,1.85$ where a saturation effect might be expected.
The double dip structure in Fig.~2, which has 
been observed experimentally~\cite{schmitt},
corresponds closely to these classical turning points. Slightly 
beyond the classical turning points a negative slope represents a 
region stable against changes in the input state.
In the 60:40 case, the input-output curve in Fig.~\ref{figio} suggests that significant squeezing would occur
near $N=1.62$ where the slope changes sign. The quantum-field theoretic results discussed next do not predict
noise reduction for $N=1.62,\zeta=2\pi$ --- in sharp contrast to the simplified
classical input-output picture the quantum theory actually predicts 15 dB excess noise in this 
case~\cite{cleo97} which the truncated Wigner theory also predicts. 
A simple explanation for this disagreement lies in the 
difference between a loop and an interferometer
with only one nonlinear pathway. 
In the former situation quantum theory allows 
quantum phase diffusion to develop 
in both arms independently whereas the 
classical argument would have input noise appear 
reduced in strength in both arms.
Clearly this is exacerbated in the 60:40 loop compared to a 90:10 loop.
The heuristic argument is  expected 
to disagree with quantum theory for a 90:10 loop
but with a much reduced error compared to the 60:40 case.
Quantum field-theoretic results for the 90:10 case 
will now be discussed in more detail.

For a fiber loop wih a beamsplitter transmission of 90\% 
large photon-number squeezing in the transmitted 
field (i.e., the "dark" port for a 50:50 beamsplitter arrangement) is predicted.
Large noise reduction occurs over a wide variation of coupling ratios as well.
Although optimal parameters for the largest reduction in photon-number fluctuations are still
under investigation, it is predicted 
that $11\pm1$ dB squeezing at $\zeta=3$ is possible for coherent $N\mbox{sech}(\tau)$ input pulses 
in the absence of Raman noise for $N=1.5$. Obviously smaller energy pulses can be used in combination with 
longer propagation distances. Significant reduction in the momentum fluctuations, up to 6 dB below
that for a coherent pulse, also occurs but the latter result is preliminary.
In the $N=1.5$ case the lower pulse energy arm of the Sagnac loop is only dispersive radiation
while the other arm contains a solitonic field (emergent soliton plus dispersive radiation)
whose internal noise structure was recently described by us~\cite{wernerf97}.
As discussed earlier the nonlinearity in the lower energy arm is not required to observe 
noise reduction but must be included to correctly predict the noise levels expected from a loop.

The predicted variation of the photon-number squeezing in dB versus
the input pulse amplitude parameter $N$ with $\phi(0,\tau)=N\mbox{sech}(\tau)$ coherent inputs
after propagating in a Sagnac loop of length $\zeta=\pi$ 
using the ideal QNLSE is given in Fig.~\ref{figqnlse}.
The influence of the Raman effect for $t_0=0.1$ps at room temperature (shown in Fig.~3)
was determined to be not large in this case even though the input pulse to the 
fiber had more than twice the energy of a fundamental soliton in the case of $N=1.5$.
The photon-number fluctuations are however still increased by the Raman effect
to $8.3\pm 0.4$dB below shot-noise at $\zeta=\pi$ for $N=1.5$ including 0.1dB/km losses.
There is a strong similarity
between variation with energy for a fixed distance and variation with distance for a fixed input energy since
the nonlinearity allows soliton pulses of larger(smaller) energy to experience 
similar effects over shorter(longer) distances. Both in Fig.~2 and Fig.~3
the loop output exhibits excess noise, which was also predicted~\cite{werner} and observed~\cite{friberg} 
for direct spectral filtering, and is clearly a general feature in nonlinear systems.
After $N=1.5$, the low energy pulse switches from dispersive radiation to a solitonic field with an internal
noise structure whose spectral correlations can lead to excess noise or squeezing depending
on the bandwidth of any filtering mechanism~\cite{werner,friberg}.
The noise reduction predicted for a one soliton period fiber is also given to demonstrate
the possibility of significant squeezing in short fibers provided that the pulse launched into the
fundamental propagating mode of the fiber matches the initial conditions assumed here.
The variation for $\zeta=\pi/2$ appears similar to $\zeta=\pi$ 
but with higher energy required as expected for the nonlinear fiber.
For comparison a much longer propagation distance of 16 soliton periods using $N^2=10/9$ is given in Fig.~3
so that a fundamental soliton propagates in one arm and the weak pulse experiences free propagation (non-loop case).
While solitons are not necessary the squeezing predicted for pulses in the normal dispersion regime
is less and at higher energy than soliton pulses as expected.
In this case the pulse quickly temporally broadens from the 
group velocity dispersion while the pulse spectrum broadens
initially and then reaches an equilibrium even for $N>1$. This is in contrast to the 
breathing in the anomalous regime.
In Fig.~3 the case $N=3$ is given without Raman and reaches $6.4 \pm 0.1$dB below shot noise.

In summary, photon number fluctuations for the two coupling regimes of the Sagnac loop -- slightly 
asymmetric and highly asymmetric -- have quite different characteristics.
The nonlinear optical loop mirror (slight asymmetry) usually increases
the photon-number fluctuations above the shot-noise level for 
coherent inputs~\cite{cleo97}. 
Sub-shot noise statistics is possible however the 60:40 loop is much more sensitive
and without precise control would almost certainly produce excess noise at the output.
On the other hand, photon-number noise may be significantly
reduced below the  shot-noise level easily for
highly asymmetric coupling. 
We have shown for the first time that large photon-number squeezing
of solitonic fields can be produced by a Sagnac loop over a significant
range of input energies and the squeezing is larger
than predicted for spectrally filtered optical fiber solitons~\cite{wernerf97}. 
Despite this  similarities in the output photon statistics for the two approaches  exist
due to the use of the same nonlinear propagation to produce the internal quantum correlations
of a quantum soliton not present in the initial coherent state. 
In essence a nonlinear interferometer is capable of producing optical pulses
closer to true quantum solitons than their coherent counterparts from mode-locked lasers.
The application of these ideas to controlling fluctuations in atomic wavepackets is particularly
interesting given recent experimental progress with Na atom coherent condensates with
densities $\approx 10^{15}$ cm$^{-3}$ using optical traps~\cite{opticaltrap}. 
This is several orders of magnitude larger than estimated
to observe atomic solitons~\cite{zhangws:94}.

\begin{figure}
\caption{Transmitted pulse photon-number scaled in 
units of $\bar n$ versus the input amplitude $N$ 
with $\phi(0,\tau)=N\mbox{sech}(\tau)$ inputs.
The input-output curve is shown for $\zeta=\pi,2\pi$ using  a 90:10 loop 
and  also for $\zeta=2\pi$ for a 60:40 loop.
\label{figio}}
\end{figure}
\begin{figure}
\caption{Photon-number variance (dB) at $\zeta=\pi/2,\zeta=\pi$ 
for 90:10 Sagnac loop with $\phi(0,\tau)=N\mbox{sech}(\tau)$ coherent inputs using the ideal
QNLSE.\label{figqnlse}}
\end{figure}
\begin{figure}
\caption{Photon-number variance (dB) versus propagation distance for a coherent $\phi(0,\tau)=N\mbox{sech}(\tau)$ pulse
inputs  into a 90:10 interferometer. The horizontal axis is rescaled for the case $N^2=10/9$ (solid line) 
where the lower energy pulse undergoes free propagation with no Raman. Sagnac loop results are given for
$N=1.5$(dashed line) without Raman  and  with Raman
using $t_0=0.1$ps ($\bar n=10^9$) at phonon temperature of 300K. 
An example of a non-solitonic input,$k^{''}>0$, is shown for $N=3$ without Raman. $N^2=10/9$ and $N=3$
cases used truncated Wigner theory averaged over 5000 trajectories.
\label{fig3}}
\end{figure}
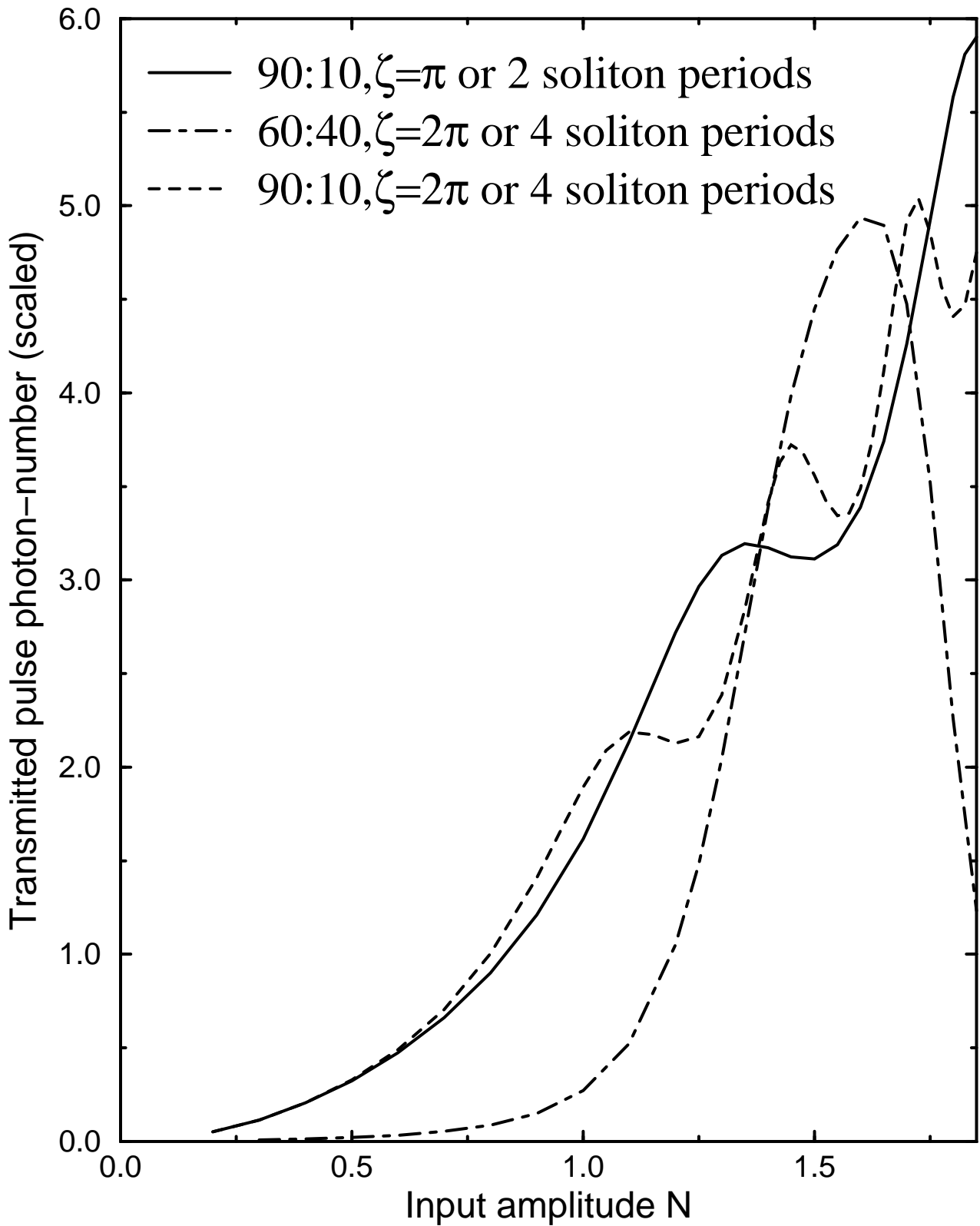
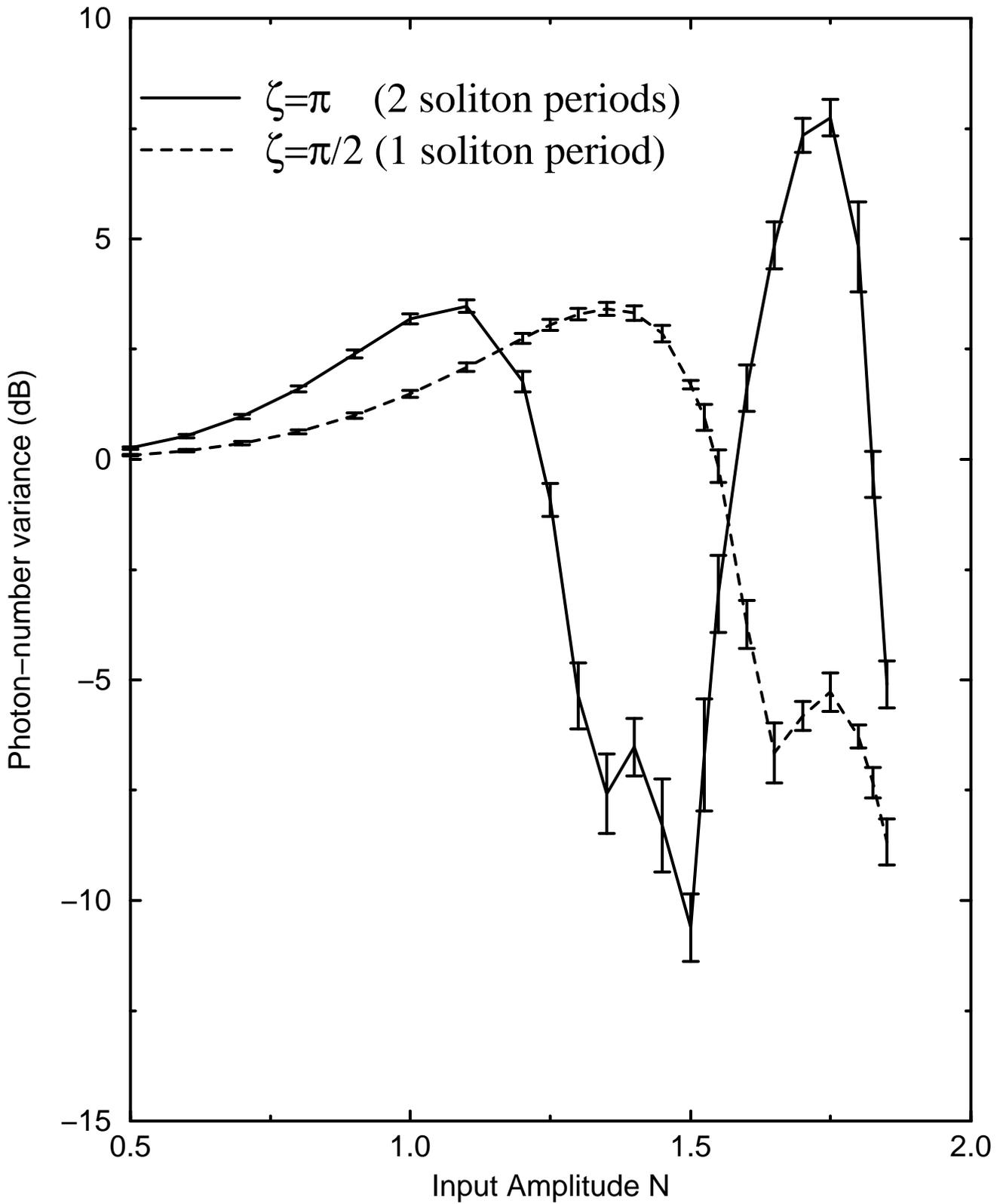
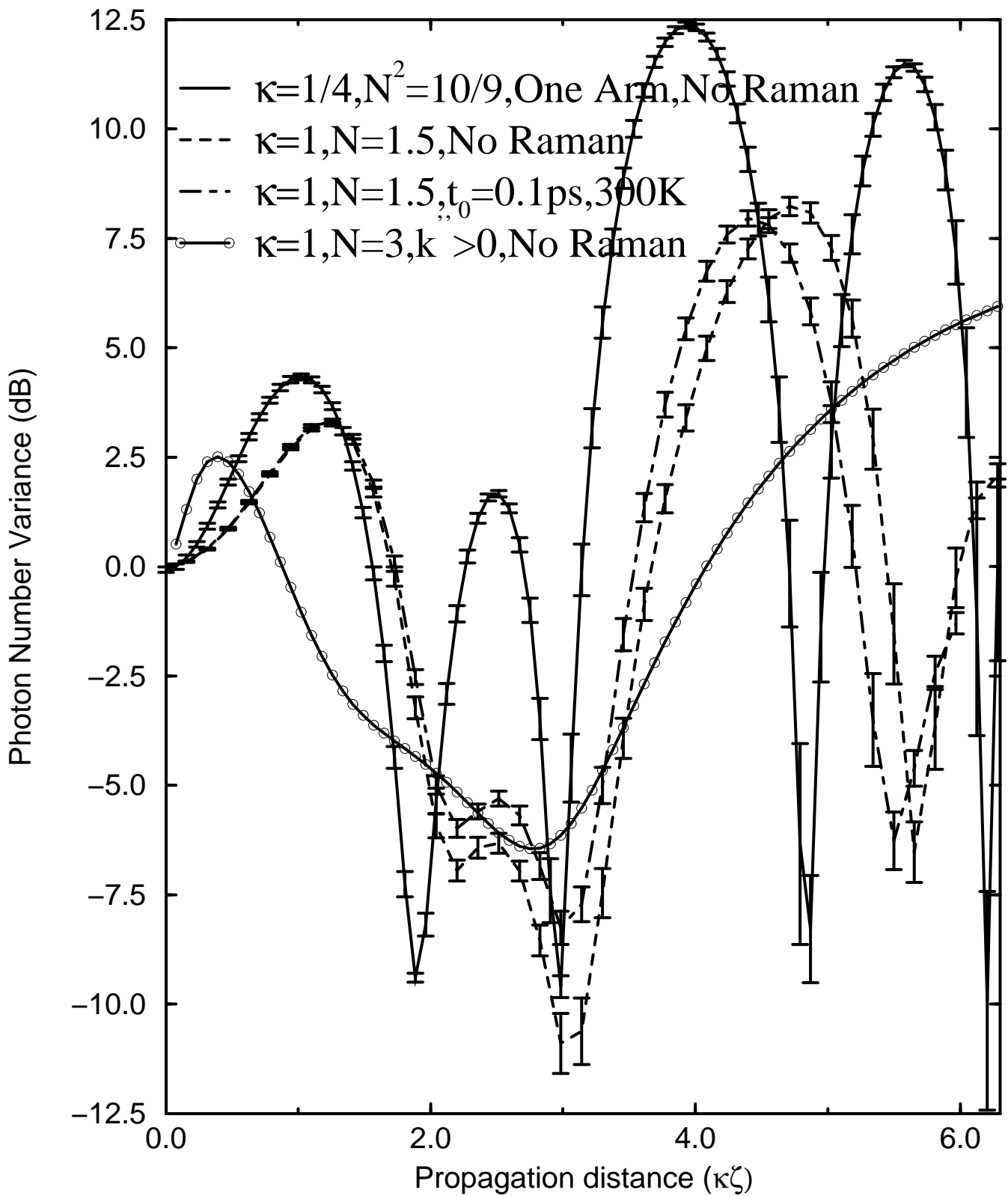
\end{document}